\documentclass{elsart3p}

 \usepackage{graphicx}

\begin{document}

\begin{frontmatter}

\title{Spin Vortex in Magnon BEC of Superfluid $^3$He-B}
\author{Yu.M. Bunkov$^1$  and G.E. Volovik$^{2,3}$}
\address{1: MCBT, Institute Neel, CNRS/UJF, Grenoble, 38042, France\\
2: Low Temperature Laboratory, Helsinki University of
Technology, Finland
\\3: Landau Institute for Theoretical Physics, Russia}


\received{12 June 2005}



\begin{abstract}

The phenomenon of the spontaneous phase-coherent precession of
magnetization in superfluid $^3$He and the related effects of spin
superfluidity are based on the true Bose-Einstein condensation of
magnons. Several different magnon BEC states have been observed:
homogeneously precessing domain (HPD); BEC condensation in the
spin-orbit potential trap ($Q$-balls); coherent precession with
fractional magnetization; and two modes of the coherent precession
in squeezed aerogel.  The spin superfluidity effects, like spin
Josephson phenomena, spin current vortices, spin phase slippage,
long distance magnetization transport by spin supercurrents have been
observed.

\end{abstract}


\end{frontmatter}


Bose-Einstein condensation (BEC) is a  phenomenon of  formation of
collective quantum state, in which the macroscopic number of
particles is governed by a single wave function. The phenomenon of
Bose-Einstein condensate was predicted by Einstein in 1925.  For a
review see, for example Ref. \cite{revBEC}. The almost perfect BEC
state was observed in ultra could atomic gases. In Bose liquids, the
BEC is strongly modified by interactions, but still remains the key
mechanism for the formation of a coherent quantum state in Bose
systems, which exhibits the phenomenon of superfluidity
characterized by  non-dissipative superfluid mass currents
discovered first in $^4$He by P.L. Kapitza \cite{Kapitza}.
Superfluidity proved to be a more general phenomenon: superfluid
mass current has been found in Fermi liquid $^3$He;  superfluidity
of electric charge -- superconductivity --  is known in metals;
superfluidity of chiral charge is discussed in quantum chromodynamics; color superfluidity -- in quark matter and
baryonic superfluidity -- in neutron stars; etc. Here we discuss magnon BEC with spin current superfluidity.

Strictly speaking, the theory of superfluidity and Bose-Einstein
condensation is applicable to systems with conserved $U(1)$ charge
or particle number. However, it can be extended to systems with a
weakly violated conservation:  it can be
applicable to a system of sufficiently long-lived quasiparticles --
discrete quanta of energy that can be treated as condensed matter
counterpart of elementary  particles. In magnetically ordered
materials, the corresponding propagating excitations are magnons --
quanta of spin waves. Under stationary conditions the density of
thermal magnons is small, but they can be pumped  by resonance
radio-frequency (RF) field (magnetic resonance). One may expect that
at very low temperatures, the non-equilibrium gas of magnons could
live a relatively long time, sufficient for formation of coherent
magnon condensate.

Recently there appeared a number of articles, where authors claimed
the observation of BEC of quasiparticles: excitons \cite{exit} and
magnons \cite{Democ}. To claim the
observation of BEC one should demonstrate the spontaneous emergence
of coherence  \cite{Snoke},  and, even better, to show  interference between two
condensates. Since the spontaneous coherence has not been observed
directly, the observation of BEC in the above articles is
still under question.

In superfluid $^3$He-B, the formation of a coherent state of
magnons was discovered about 20 years ago \cite{HPD}. In  pulsed NMR
experiments  the spontaneous formation of a domain with fully
phase-coherent BEC of magnons has been observed even in the presence
of inhomogeneous magnetic field. The main feature of this Homogeneously 
Precessing Domain (HPD)  is
the induction decay signal, which rings in many orders of magnitude
longer, than prescribed by inhomogeneity of magnetic field. This
means that  spins precess NOT with a local Larmor frequency, but
precess coherently with  common frequency and phase. This BEC can
be also created and stabilized by continuous NMR pumping. In this
case the NMR frequency plays a role of magnon chemical potential,
which determines the density of magnon condensate. The interference
between two condensates has also been demonstrated.
 It was shown that HPD exhibits all the properties of spin superfluidity (see Reviews
\cite{FominLT19,BunkovHPDReview}). The main property is the
existence of spin supercurrent,  which transports the magnetization
on a macroscopic distance more than 1 cm long. This spin
supercurrent  flows separately from the mass current, in contrast to
 the spin-polarized  $^3$He-A$_1$, where
spin is transported by the mass current. Also the  related phenomena
have been observed: spin current Josephson effect; phase-slip
processes at the critical current; and spin current vortex  --  a
topological defect which is the analog of a quantized vortex in
superfluids and of an Abrikosov vortex in superconductors; etc.

The spin-orbit coupling  due to dipole-dipole interaction between the
spins of $^3$He atoms is responsible for the interaction
between magnons. This interaction is relatively small, as a result  HPD
represents almost pure BEC of magnons \cite{V,B}. In typical $^3$He
experiments, the critical temperature of magnon condensation is 3
orders of magnitude higher, than the temperature of superfluid
transition; i.e.  magnons undergo the condensation as soon as
chemical potential and spin-orbit coupling allow for this process.
The superfluid $^3$He is a very unique complex macroscopic quantum
system with broken spin, orbital and gauge symmetries, where the
structure of spin-orbit coupling  can be varied experimentally. This 
leads to experimental realization of different types of magnon BEC
 in $^3$He-B \cite{HPD2};  non-topological solitons  in $^3$He-B called
$Q$-balls in high energy  physics \cite{Qbal}; and also  magnon BEC
in $^3$He-A \cite{JapCQS}.


As distinct from the static equilibrium magnetic states with broken
symmetry,  the phase-coherent  precession is the dynamical state
which experiences the off-diagonal long-range order:
\begin{equation}
\left<\hat S_+\right>=S_+ =S\sin\beta e^{i\omega t+i\alpha} ~.
\label{precession}
\end{equation}
Here $\hat S_+$ is the operator of spin creation; $S_+=S_x+iS_y$;
${\bf S}=(S_x,S_y,S_z=S\cos\beta)$ is the vector of spin density
precessing in the applied magnetic  field ${\bf H}=H\hat{\bf z}$;
$\beta$, $\omega$ and $\alpha$ are correspondingly the tipping
angle, frequency and  the phase of precession. In the  modes under
discussion here, the magnitude of the precessing spin $S$ equals to an
equilibrium value of spin density $S=\chi H/\gamma$ in the applied
field, where $\chi$ is spin susceptibility of $^3$He-B, and $\gamma$
the gyromagnetic ratio of the $^3$He atom (the coherent state with
half of magnetization $S=(1/2)\chi H/\gamma$ has been also observed
in bulk $^3$He-B \cite{half}). Similar to the conventional mass
superfluidity which also experiences the off-diagonal long-range
order, the spin precession in Eq.(\ref{precession}) can be rewritten
in terms of the complex scalar order parameter \cite{FominLT19,V,B}
\begin{equation}
\left<\hat\Psi\right>=  \Psi=\sqrt{2S/\hbar}\sin
\frac{\beta}{2}~e^{i\omega t+i\alpha}~,
 \label{OrderParameter}
 \end{equation}
If the spin-orbit interaction is small and its contribution to the
spectrum  of magnons  is neglected in the main approximation (as it
typically occurs in $^3$He), then $\hat\Psi$ coincides with the
operator of the annihilation of magnons, with the  number density of
magnons being equal to condensate density:
\begin{equation}
 n_M=\left<\hat\Psi^\dagger\hat\Psi\right>= \vert \Psi\vert^2 =\frac{S-S_z}{\hbar}~.
\label{NumberDensity}
\end{equation}
This implies that the precessing states in superfluid $^3$He realize
the almost complete BEC of magnons. Spin-orbit coupling
produces a weak interaction between magnons and leads to the
interaction term in corresponding Gross-Pitaevskii equation for the
BEC of magnons (further we use units with $\hbar=1$):
  \begin{eqnarray}
 \frac{\delta F}{\delta \Psi^*}=0~,
  \label{GP}
  \\
 F=\int d^3r\left(\frac{\vert\nabla\Psi\vert^2}{2m_M} -\mu\vert\Psi\vert^2+{\bar E}_D(\vert\Psi\vert^2)\right),
\label{GL}
\end{eqnarray}
Here the role of the chemical potential $\mu=\omega-\omega_L$ is
played by the shift of the precession frequency from  the Larmor
value $\omega_L=\gamma H$. In coherent states,
the  precession frequency $\omega$ is the same throughout the whole
sample even in  the nonuniform field; it is determined by the number
of magnons in BEC,  $N_M=\int d^3r n_M$, which is conserved quantity
if  the dipole interaction is neglected. In the regime of continuous
NMR,   $\omega$ is the frequency of the applied RF field,
$\omega=\omega_{\rm RF}$, and the chemical potential
$\mu=\omega_{\rm RF}-\omega_L$ determines the magnon density.
Finally, $m_M$ is the magnon mass; and ${\bar E}_D$ the  dipole
interaction averaged over the fast precession. The general form of
${\bar E}_D(\vert\Psi\vert^2)$  depends on the orientation of the
orbital degrees of freedom described by the unit vector $\hat{\bf
l}$ of the orbital momentum, see Ref.  \cite{BV}.


In the coherent precession in bulk $^3$He-B  the  spin-orbit
coupling orients  $\hat{\bf l}$ along $ {\bf H}$. In this case the interaction
term ${\bar E}_D$ 
has the form different from conventional 4-th order term in dilute
gases \cite{V}:
 \begin{eqnarray}
 {\bar E}_D=0~,~\vert\Psi\vert^2<\frac{5}{4}  S~,
\label{0}
 \\
{\bar E}_D=\frac{8}{15}\chi\Omega_L^2
\left(\frac{\vert\Psi\vert^2}{S}-\frac{5}{4}
\right)^2~,~\vert\Psi\vert^2>\frac{5}{4}     S~.
     \label{FHPD}
  \end{eqnarray}
Here $\Omega_L\ll \omega_L$ is the Leggett frequency which
characterizes the dipole interaction. If the chemical potential
$\mu$ is negative, i.e. $\omega<\omega_L$, the
minimum of the Ginzburg-Landau (GL) energy ${\bar
E}_D(\vert\Psi\vert^2) -\mu\vert\Psi\vert^2$ corresponds to
$\Psi=0$, i.e. to the static  state with  equilibrium
magnetization ($\beta=0$).  For $\mu>0$ ($\omega>\omega_L$)  
the minimum of the GL energy
corresponds to $\vert \Psi\vert^2/S = (5/4) + (15/32)\tilde \mu$,
where $\tilde \mu =(\omega^2-\omega_L^2)/\Omega_L^2$.
The consequence of such a peculiar profile of the interaction term is
that  as distinct from the dilute gases the formation of the magnon
BEC starts with the finite magnitude $\vert\Psi\vert^2=(5/4)S$. This
means that the coherent precession starts with a tipping angle equal
to the magic Leggett angle, $\beta= 104^\circ$, and then the tipping
angle increases with increasing frequency shift. This coherent state
called the HPD   persists  indefinitely, if one applies a small RF
field to compensate the losses of magnons caused by small spin-orbit
interaction.

In conventional magnetic systems,  magnetization precesses in the
local field with the local frequency shift and thus experiences
dephasing in the inhomogeneous field. In the case of  magnon BEC,
the rigidity of the order parameter (the gradient term in
Eq.~\ref{GL})  plays an important role. The spatial dephasing leads
to the gradient of chemical potential. This in turn excites the spin
supercurrents, which finally  equilibrate the chemical potential. In
the steady state of magnon BEC the gradient of  a local field is
compensated by small gradient of magnon density $\vert \Psi \vert
^2$ in such a way, that $\omega$ and $\alpha$
remain homogeneous throughout the whole sample.

\begin{figure}
\begin{center}
\includegraphics[%
  width=1.1\linewidth,
  keepaspectratio]{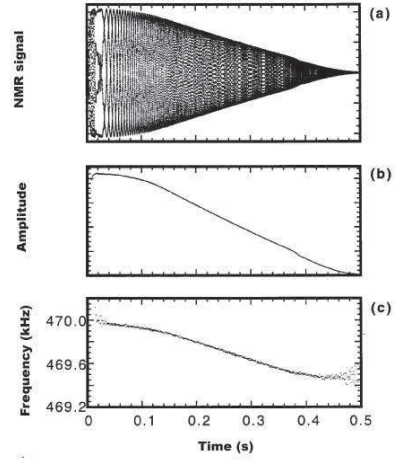}
\end{center}
\caption{(Color online) The typical signal of induction decay from
the BEC of magnons; (a) stroboscopic record of the signal; (b)
amplitude of signal; (c) frequency of the signal. Frequencies 469.95
and 469.4 kHz correspond to  Larmor frequency at the top and the
bottom of the cell. } \label{hpd2}
\end{figure}

In a pulsed NMR experiment,  magnetization is deflected by a
strong  RF pulse. The typical induction signal after the pulse in
the cell with  a large gradient of magnetic field along the axis of
the cell is shown in Fig.~\ref{hpd2}. Due to the field gradient the
induction signal should dephase and disappear in about 1 ms.
Instead, after a transient process of about 2 ms, the induction
signal acquires an amplitude corresponding to a 100\% coherent
precession of the deflected magnetization with the spontaneously
emerging phase. This coherent state lives 500 times longer than the
dephasing time caused by inhomogeneity.

\begin{figure}
\begin{center}
\includegraphics[%
  width=1.1\linewidth,
  keepaspectratio]{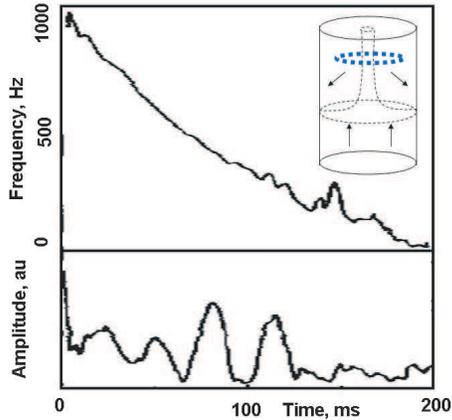}
\end{center}
\caption{NMR signature of the spin vortex in magnon BEC (HPD). The frequency decays with time in the 
same way as in conventional vortex-free precession, while the amplitude of the HPD signal is nearly zero because of compensation of signals from opposite sides of the cells where the phase $\alpha$ of precession differs by $\pi$. } \label{vt}
\end{figure}

What happens with the magnon system during the transient period?  As
soon, as at the higher field end of the cell the deflection of
magnetization becomes smaller than 104$^\circ$, the spin-orbit
interaction cannot anymore compensate the gradient of magnetic
field. The cell splits into two domains, in one of them the
magnetization is stationary, while in the other one magnons condense
with the density  close to $\vert \Psi\vert^2 = (5/4)S$, i.e. the
magnetization precesses with $\beta$ slightly above  104$^\circ$. In
the subsequent process of relaxation caused by the non-conservation
of magnon number,  the volume of the BEC condensate (HPD) decreases.
During the relaxation, the BEC  does not loose the phase coherence,
but its chemical potential (precession frequency) changes. The
frequency of HPD corresponds to Larmor frequency at the boundary of
the domain  and slowly changes in time with relaxation as the boundary moves
down (see Fig.~\ref{hpd2}; the amplitude of the signal exactly
corresponds to the record of the frequency).

The small RF field can compensate the HPD relaxation by pumping
additional magnons. In this steady state many 
phenomena of spin superfluidity have been observed: Josephson effect; magnetization
transport by spin supercurrent;  critical spin current velocity,
phase slip; etc. Here we will discuss the observation of a spin vortex \cite{Vort}.

 In HPD with a single vortex in the central part of  cylindrical cell  the
phase $\alpha$ of precession changes by $2\pi$  around the center, i.e.
 $\alpha$ is opposite on the opposite sides of the cell. In the
central part of the cell, i.e. in the vortex core,  the magnetization remains vertical and does not
precess. We  created  HPD with a spin vortex by applying the quadrupole RF field.
For this purpose we connected two parts of the saddle NMR coil in opposite
directions, so that the phase of RF field (and consequently the phase $\alpha$) was opposite at the opposite sides of the cell. By these NMR coils we observed practically the same HPD
signal as in the conventional arrangement with the parallel connection of the coils, though with a slightly reduced  amplitude.Ê This shows that we created  HPD with opposite $\alpha$ on opposite sides of the cell. To verify this we  installed a pair of small pick-up coils at the top of the cell connected in usual way. When we switched off the RF field, the pick up coils received a very small RF signal from HPD, while the
frequency of this signal corresponded to the full  HPD signal. This
means that HPD generated the signal with the opposite phase at the two sides of the pickup coils, which nearly compensated each other (see
Fig.~\ref{vt}).
This corresponds to  HPD with a circular gradient of $\alpha$, as shown in inset of Fig.~\ref{vt}. The magnetization is oriented vertically in the vortex core. On theÊ 
periphery of the cell it precesses with tipping angle 104$^\circ$,
and with $2\pi$ phase winding around the center. This type of HPD should
radiate at frequency, which corresponds to the Larmor field on the
boundary of HPD, but should not produce any signal in the pick-up coil.
However,Ê a small signal appears due to asymmetry of the pick-up coil;  oscillations of this signal may correspond toÊ nutations of the vortex
core.

Of course, the HPD has been observed, studied and explained on the
basis of theory of spin superfluidity and non-linear NMR long time
ago \cite{HPD}. However, the consideration of this phenomenon in
terms of magnon BEC not only demonstrates the real system with
the BEC of excitations, but also allows us to simplify the problem
and to study and search for the other types of magnon BEC in
$^3$He, such as $Q$-balls \cite{Qbal}; HPD$_2$ state found in a
deformed aerogel \cite{HPD2}; coherent precession in $^3$He-A also
found in the deformed aerogel \cite{JapCQS}; etc.

\end{document}